\documentclass[%
aip,
jmp,%
amsmath,amssymb,
preprint,%
author-numerical,%
]{revtex4-1}
\usepackage{graphicx}
\usepackage{amsmath}
\usepackage{epsfig}

\begingroup
\renewcommand{\section}[2]{}

\usepackage{color}

\begin{document}

\title[]{Simulated evolution of three-dimensional magnetic nulls leading to generation of cylindrically-shaped current sheets}

\author{Sanjay Kumar and R. Bhattacharyya}
\affiliation{ Udaipur Solar Observatory, Physical Research Laboratory, Dewali, Bari Road, Udaipur-313001, India}

\begin{abstract}
The performed magnetohydrodynamic simulation examines  the importance of magnetofluid evolution
which naturally leads to current sheets in the presence of three-dimensional (3D) magnetic  nulls.  Initial magnetic 
field is constructed by superposing a 3D force-free field on a constant axial magnetic field. 
The initial field supports  3D magnetic nulls having classical spine axis and dome-shaped fan surface; and exerts non-zero Lorentz force on 
the magnetofluid. Importantly, the simulation identifies the development of current sheets near 
the 3D magnetic nulls. The morphology of the current sheets is similar to a cylindrical surface
where the surface encloses the spine axis. The development is because of favorable deformation of 
magnetic field lines constituting the dome-shaped fan surface.  The deformation of field lines 
is found to be caused from the flow generated by magnetic reconections at current sheets which 
are located away from the cylindrically-shaped current sheets. 
\end{abstract}


\date{\today}
\maketitle

Solar flares are explosive  phenomena observed in the magnetically dominated solar atmosphere. 
Standardly, the magnetic energy stored in the twisted coronal magnetic field lines (MFLs) is believed to drive 
flares  while getting converted into heat and kinetic energy of mass motion through the process
of magnetic reconnection (MR) {\cite{shibata2011, priest-new}}.

To understand the onset of MRs, notable is the non-diffusive nature of the coronal 
magnetofluid, stemmed from large magnetic Reynolds number ($R_M = v L/\lambda$, in usual notations) {\cite{aschwanden}}.
Consequently, the MFLs remain tied to fluid parcels---a condition referred to as the ``flux-freezing" {\cite{priest-new}}. 
In contrast, MR being a diffusive process demands the generation of small length scales in magnetic field, or equivalently, current 
sheets (CSs)---the sites with intense current density ${\bf{J}}=\nabla\times{\bf{B}}$---to locally reduce 
$R_M$. The spontaneous origin of CSs is ensured by the Parker's magnetostatic 
theorem {\cite{parker-book, parker-1972, parker-2012}} according to which, a magnetofluid
with infinite electrical conductivity and complex magnetic topology is unable to 
maintain an equilibrium having continuous magnetic field. Translation of the theorem to an 
evolving magnetofluid then favors onset of CSs when the magnetofluid relaxes toward an 
equilibrium. With a small magnetic diffusivity, the magnetofluid becomes diffusive near CSs and 
supports MRs. Post MRs, the magnetofluid once again attains the flux-freezing and may lead to  the 
secondary CSs, initiating further MRs. Importantly, these 
spontaneous MRs---intermittent in space and time---may shape up the dynamics of a 
large $R_M$ magnetofluid {\cite{sanjay-apj}}.

In a complex three dimensional (3D) magnetic field, preferential sites of MRs
can be magnetic nulls {\cite{sanjay2016,pontin2007}} which represent the skeletal topology of the field. 
Interesting are the 3D nulls, believed to be present in abundance
in the solar corona {\cite{longcope}}. Notably, MRs at 3D nulls with their
classical spine and dome-shaped fan  structures can be important
in circular ribbon flares {\cite{messon, wang}} as well as some other eruptive phenomena \cite{urra}.
With CSs being requisite for MRs, it is then imperative  to explore evolution of MFLs near 
3D nulls toward autonomous development of CSs.
To allow for the autonomous development,  magnetohydrodynamic (MHD) simulations are required to be conducted in 
congruence with the magnetostatic theorem. For the purpose, we 
simulate viscous relaxation  of a thermally 
homogeneous,  incompressible magnetofluid with infinite electrical 
conductivity  {\cite{sanjay-pap, complexity-pap}} from a suitably constructed initial state.

The initial magnetic field ${\bf{B}}$ needs to be prescribed such as to have inherent 3D nulls along with 
non-zero Lorentz force, required for dynamical evolution. The prescribed ${\bf{B}}$ used here is 
superposition of an uniform 
field  ${\bf{B}_1}=\hat{e}_z$ with a 3D linear force force-free field  ${\bf{B}_2}=\{{B_x}^{2}, {B_y}^{2}, {B_z}^{2}\}$ where

\begin{eqnarray}
\label{comp2}
 {B_x}^2 = \sqrt{3} \sin\left( x\right) \cos \left(y
\right) \sin\left(z\right) 
 +  \cos\left( x\right) \sin\left( y\right) \cos\left( z\right),\\
\label{comph2}
 {B_y}^2 = -\sqrt{3}\cos\left( x\right)\sin\left( y\right) \sin\left(z\right) 
 + \sin\left( x \right) \cos\left( y\right) \cos\left( z\right), \\
\label{comph3}
 {B_z}^2 = 2 \sin\left( x \right) \sin\left( y\right) \sin\left(z \right).
\end{eqnarray}

\noindent ${\bf{B}_1}$ and ${\bf{B}_2}$ satisfy $\nabla\times{\bf{B}}_1=0$ and $\nabla\times{\bf{B}}_2=\sqrt{3}{\bf{B}}_2$ 
respectively. Importantly, ${\bf{B}_2}$ is known to have 3D nulls {\cite{sanjay-pap}}. With 
${\bf{B}_1}$ lacking any nulls, ${\bf{B}}$ is expected to has the identical skeletal topology as ${\bf{B}_2}$ and 
additionally, a non-zero Lorentz force. Explicitly, 

\begin{eqnarray} 
{\bf{B}}={\bf{B}_1}+d_0{\bf{B}_2}, 
\end{eqnarray}
where the constant $d_0$ relates the amplitudes of the two 
fields and determines the deviation of ${\bf{B}}$ from its force-free equilibrium.  
The components of ${\bf{B}}=\{B_x, B_y, B_z\}$ are
\begin{eqnarray}
\label{super}
{B_{x}}= d_0(\sqrt{3} \sin\left( x\right) \cos \left(y
\right) \sin\left(z\right) 
 +  \cos\left( x\right) \sin\left( y\right) \cos\left( z\right)),\\
{B_{y}}= d_0(-\sqrt{3}\cos\left( x\right)\sin\left( y\right) \sin\left(z\right) 
 + \sin\left( x \right) \cos\left( y\right) \cos\left( z\right)),\\
{B_{z}}=1+d_0(2 \sin\left( x \right) \sin\left( y\right) \sin\left(z \right)),
\end{eqnarray}
defined in an uniform triply periodic Cartesian domain of period $2\pi$.  The corresponding Lorentz force is
\begin{eqnarray}
{\bf{J}}\times{\bf{B}}=d_0(\sqrt{3}){\bf{B}}_2\times{\bf{B}}_1 ~.
\label{lorentz}
\end{eqnarray}
which is non-zero for $d_0\neq 0$. We select $d_0=0.75$ to set a  magnitude 
of the initial Lorentz force which is optimal for the simulation. 

\begin{figure}[htp]
\centering
\includegraphics[angle=0,scale=.30]{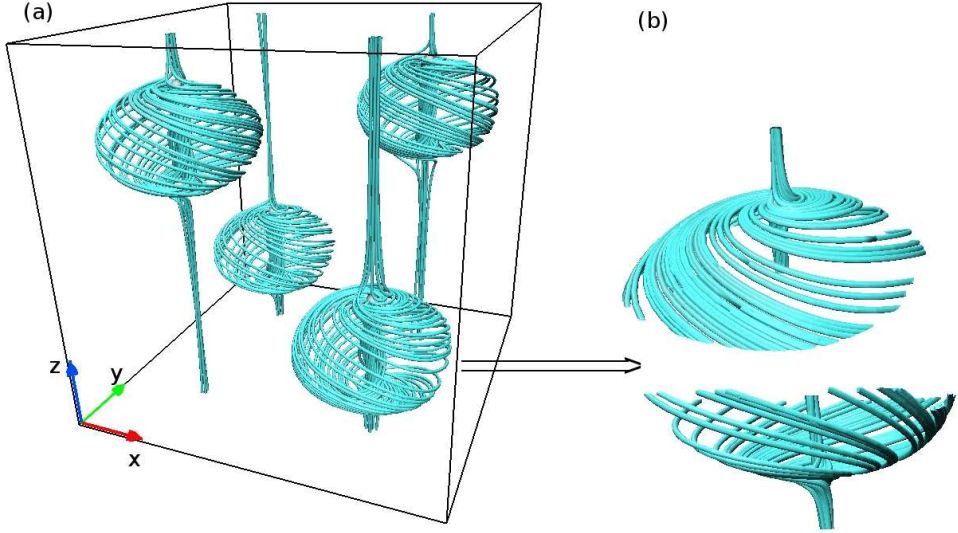}
\caption{Panel (a) demonstrates the presence of eight 3D magnetic nulls in the initial field ${\bf{B}}$. 
Noticeably, the MFLs associated with each 3D null display dome-shaped fan surface and spine axis (panel (b)).} \label{initialfield}
\end{figure}

In Figure \ref{initialfield} we plot MFLs of the  
${\bf{B}}$ overlaid with magnetic nulls. The figure visually confirms the presence of 3D nulls, eight in total,
in the form of points and a complete absence of 2D nulls. Notably, the eight 3D nulls can be 
grouped into four pairs, where each pair shares a common spine axis. For clarity, two elements of a given pair, 
with an in-between artificially created void, is shown separately in
the Figure \ref{initialfield}(b). The fan surface of each 3D null 
being dome-shaped, the resulting combined fan  of a pair is the 
visible closed  surface spanned by MFLs.

The MHD simulation is performed by using the numerical model 
EULAG-MHD \cite{smolar-2013}. Relevantly, the model advection 
scheme {\cite{Smolar06}} produces locally adaptive residual 
dissipation in response to generation of under-resolved scales in
field variables to effectively regularize these scales {\cite{sanjay2016, sanjay-apj}}, in the 
spirit of implicit large eddy simulation (ILES) subgrid-scale turbulence models {\cite{morgolin}}.
Importantly, the presented simulations reported rely on the proven 
ILES mode of EULAG-MHD in regularizing the under-resolved scales through onset of MRs, concurrent and 
collocated with developing CSs {\cite{kumar-bhattacharyya, sanjay2016}}.

\begin{figure}[htp]
\centering
\includegraphics[angle=0,scale=.50]{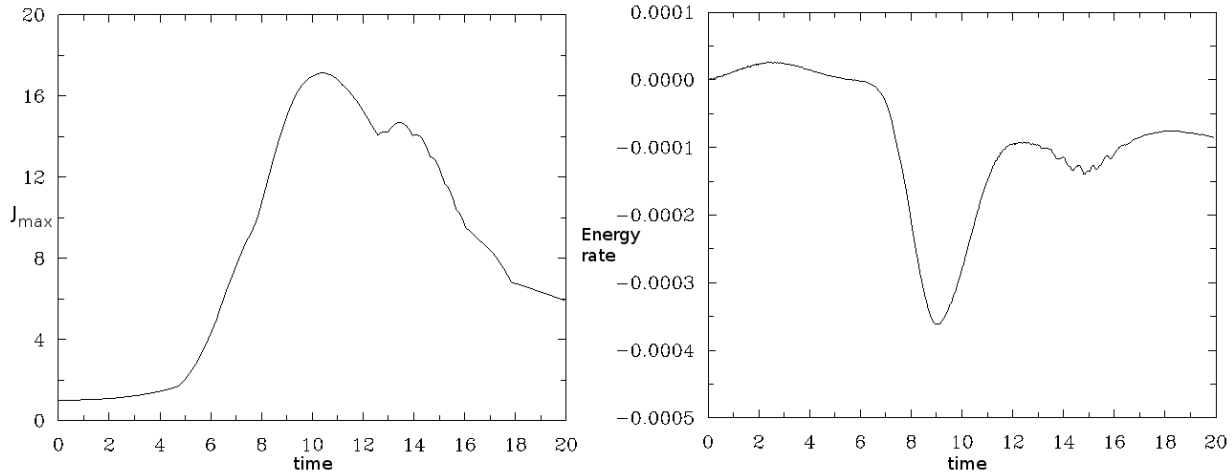}
\caption{ Panel (a) depicts the history of maximum value of current density ($\mid{\bf{J}}\mid_{\rm max}$). 
The current density is normalized with its initial value.  Evident is 
two distinguishable peaks in $\mid{\bf{J}}\mid_{\rm max}$ centered at $t\approx10$s and $t\approx 14$s. Panel (b) shows numerical magnetic energy 
rate with time, normalized to the initial total energy. The deviations in the rate 
from its analytical value zero are during the formation of the peaks in  $\mid{\bf{J}}\mid_{\rm max}$.     } \label{jmax}
\end{figure}

The computation is conducted on the $128\times 128\times 128$ uniform grid, resolving the 
computational domain $\Gamma$ spanning $[0, 2\pi]^3$.
The coefficient of viscosity and the mass density are set 
to $0.002$ and $1$ respectively. 
The initial Lorentz force pushes the magnetofluid from an initial motionless state and develops dynamics. 
To elucidate and understand the development of 
current sheets, in Figure \ref{jmax}(a) we plot the time profile of maximal volume current density 
($\mid{\bf{J}}\mid_{\rm max}$). The plot shows 
a monotonic rise in $\mid{\bf{J}}\mid_{\rm max}$ until around $t=10$s and followed by a decrease, resulting in the development 
of a peak centered at $t=10$s. In addition, a second peak is also formed at around $t= 14s$. 
The formation of such peaks in $\mid{\bf{J}}\mid_{\rm max}$ can generally be attributed to formation of CSs 
and their subsequent decay by MRs {\cite{mellor}}. To further confirm, in Figure \ref{jmax}(b), we depict numerical deviations 
of normalized magnetic energy rate from its analytical value (cf. Equation (7) of Ref. \cite{sanjay-pap}). 
After $t\approx 8$s, the curve starts deviating considerably from its analytical value of zero. 
The plot shows a maximal numerical deviation of magnitude $0.00035$ 
in the energy rate during formation of the first peak in  $\mid{\bf{J}}\mid_{\rm max}$. 
After the first peak, the model regains its desired numerical accuracy, 
losing it again by a marginal amount during formation of the
second peak in the current density (Fig. \ref{jmax}(a)). Such deviations in the energy rate 
can be attributed to the onset of  MRs. 

\begin{figure}[htp]
\centering
\includegraphics[angle=0,scale=.30]{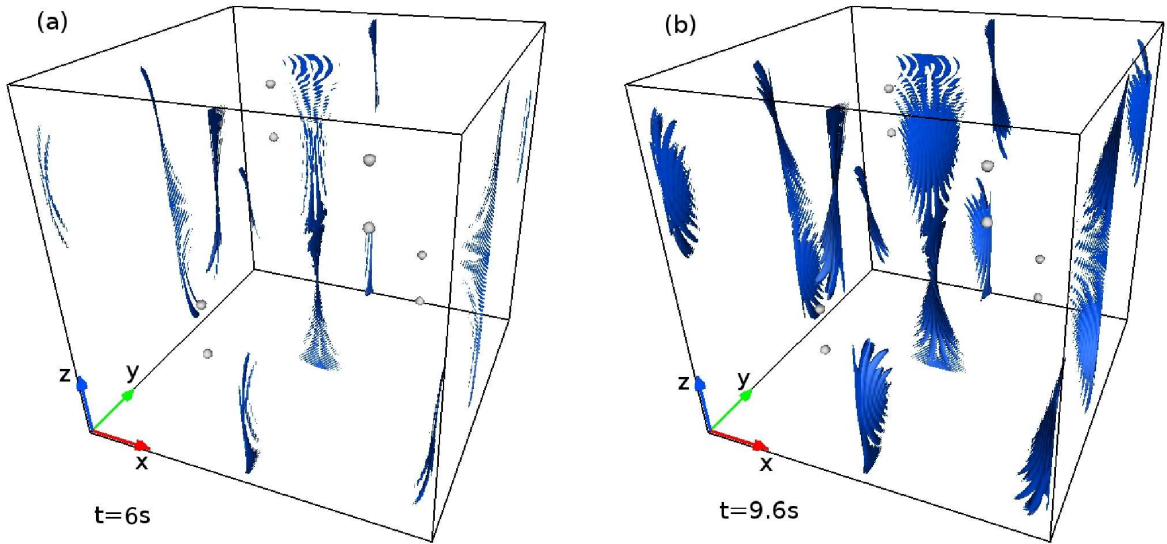}
\caption{The isosurfaces of $\mid{\bf{J}}\mid$  (in blue) having a magnitude of
$50\%$ of the $\mid{\bf{J}}\mid_{\rm max}$ ($J-50$) at $t=6$s and $t=9.6$s. The figure is overplotted 
with magnetic nulls (in grey), highlighting formation of CSs away from nulls and elongated 
along $z$. } \label{cs1}
\end{figure}

To locate CSs in the domain $\Gamma$,  we analyze the appearances of isosurface of current density $\mid{\bf{J}}\mid$ with an isovalue 
which is $50\%$ of $\mid{\bf{J}}\mid_{\rm max}$. Hereafter, we refer these isosurfaces as $J-50$ surfaces and 
identify them with CSs. Figure \ref{cs1} displays  $J-50$ surfaces (overlaid with magnetic nulls) 
at $t=6$s and $t=9.6$s.  The surfaces start appearing around $t=6$s (panel (a)) followed by its spatial 
extension with time. Notably, the surfaces are elongated along $z$ and are situated away from the 3D nulls, indicating 
the development of the elongated CSs away from the nulls. Such development of CSs can be because of
favorable contortions of contributing magnetic flux surfaces {\cite{sanjay-pap, complexity-pap}}.

\begin{figure}[htp]
\centering
\includegraphics[angle=0,scale=.30]{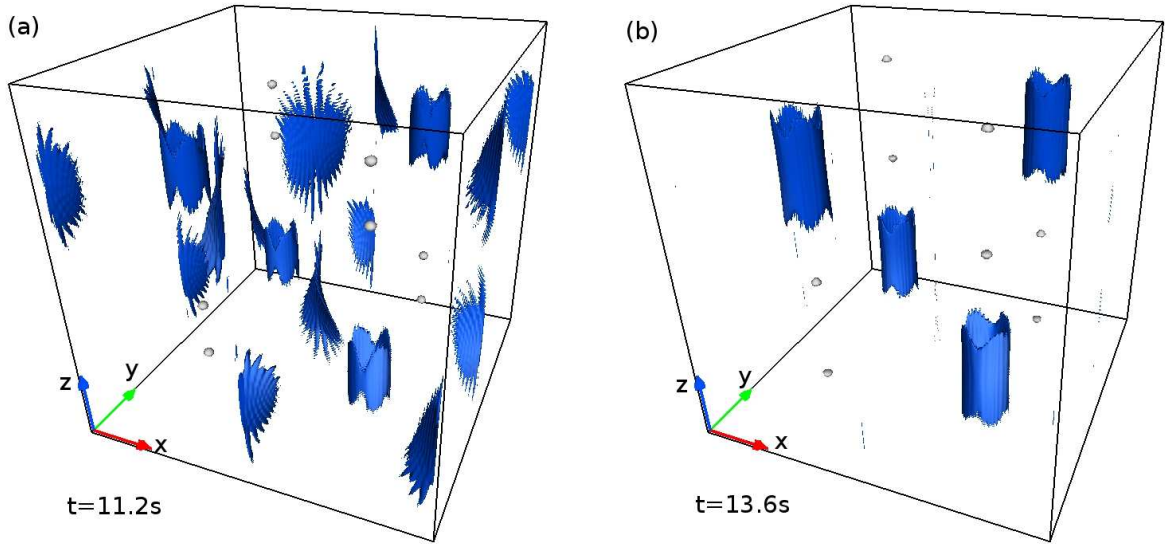}
\caption{ The $J-50$ surfaces (in blue) overlaid with magnetic nulls (in grey) at 
$t=11.2$s and $t=13.6$s. Notable is a decay of the elongated CSs along with the development of
the CSs having cylindrical geometry. } \label{cs2}
\end{figure}

In Figure {\ref{cs2}}, we show the $J-50$ surfaces at $t=11.2$s and $t=13.6$s. The figure indicates 
a constant decrease in the spatial extension of elongated $J-50$ surfaces along with the generation of  
additional $J-50$ surfaces having cylindrical geometry. This suggests the onset of the dissipation of the 
elongated CSs via MRs and a concurrent development of cylindrically-shaped CSs. 
    
\begin{figure}[htp]
\centering
\includegraphics[angle=0,scale=.25]{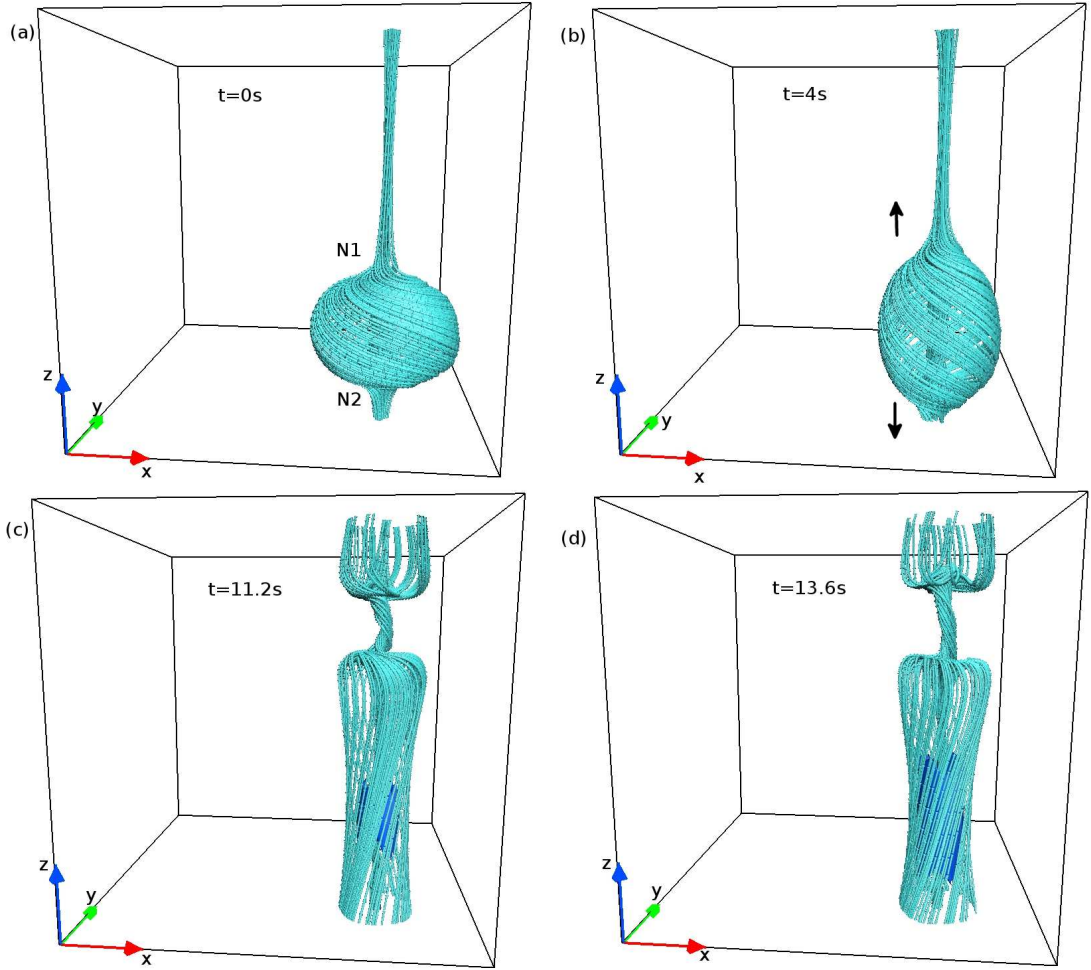}
\caption{Time evolution of MFLs (overlaid with $J-50$ surfaces) in the vicinity of a pair of 
3D nulls (marked by N1 and N2 in panel (a)). Initial motion of the nulls are marked by arrows in the panel (b).
Important is the development of the 
cylindrically-shaped  CSs by the collapse of dome-shaped MFLs structure  
toward the spine of the 3D null (marked by N1). } \label{csnull2}
\end{figure}

\begin{figure}[htp]
\centering
\includegraphics[angle=0,scale=.35]{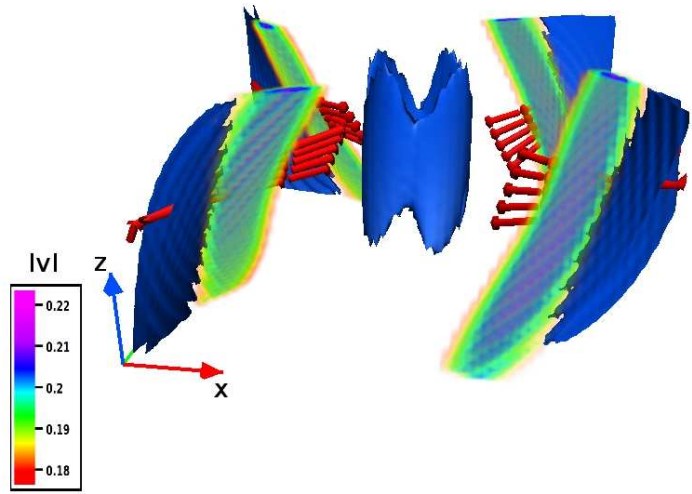}
\caption{Direct volume rendering of the magnitude of velocity field (overplotted with $J-50$ surfaces) 
generated by MRs near elongated CSs at $t=11.2$s.  The CSs are situated near the location where  
a cylindrically-shaped CS (depicted in Fig. \ref{csnull2}) develops. 
The figure is further overlaid with the streamlines of velocity field, suggesting the flow to be favorable to 
enable the collapse of the dome-shaped MFLs of the null and consequent development of CS. } \label{flow}
\end{figure}

To understand the development of  cylindrically-shaped CSs, Figure {\ref{csnull2}} plots
evolution of MFLs in the vicinity of a pair of  3D nulls (marked by N1 and N2). Under initial forcing, the nulls get shifted from their
initial locations and move in a direction marked by the arrows in panel (b). As a result, 
one of the null (marked by N2) is pushed outside the $\Gamma$ through boundary $z=0$ and, because of the 
periodicity, the identical null appears back in the $\Gamma$ from boundary $z=2\pi$. More importantly, the 
subsequent dynamics leads to the generation of flow by MRs near the elongated CSs (cf. Figure {\ref{flow}}). The 
flow symmetrically pushes the dome structure of the 3D null (marked by N1 in Fig. \ref{csnull2}) which leads to 
a favorable collapse of the structure toward the spine axis. The 
collapse enables the constituent non-parallel MFLs to come in close proximity, resulting 
in the development of a cylindrically-shaped CS around the spine  (Fig. \ref{csnull2}(c) and (d)). The subsequent dynamics 
leads to the dissipation of the CS through MR (not shown).

Overall, the performed computation investigates the spontaneous development of 
CSs in a given initial magnetic field which supports 3D nulls with MFLs structure resembling a dome. 
Importantly, the simulation demonstrates  development 
of cylindrically-shaped CSs via a favorable collapse of the dome toward the spine axis. 
Furthermore, the collapse of the dome is found to 
causally connected and preceded by reconnection of elongated CSs located elsewhere. 
Remarkably, the cylindrically-shaped CSs are geometrically similar 
to circular flare ribbons,  complete with intermittent MRs
in the neighborhood {\cite{messon, wang}}.  The similarity opens up a possibility which indicates MRs at 
the autonomously developed CSs may be crucial for the onset of circular flare ribbons. 
However, with the idealistic initial conditions used for the performed simulation, 
the documented similarity is not  decisive and warrants further attention in terms of future 
explorations with more realistic initial magnetic field.  \\

The simulations are performed using the 100 TF cluster Vikram-100 at Physical Research Laboratory, India. 
We acknowledge the visualisation software VAPOR (www.vapor.ucar.edu), for generating relevant graphics. 
The authors also thank an anonymous referee for providing valuable insights and suggestions to make the paper 
more accurate and readable. \\\\

\end{document}